\documentclass[twocolumn,showpacs,preprintnumbers,amsmath,amssymb]{revtex4}
\input psfig.sty

\usepackage{graphicx}
\usepackage{dcolumn}
\usepackage{bm}

\begin{document}


\title{Some Observational Consequences of Brane World Cosmologies}

\author{J. S. Alcaniz} \email{alcaniz@astro.washington.edu}
\affiliation{Astronomy Department, University of Washington, Seattle,
Washington, 98195-1580, USA
}%

\date{\today}

\begin{abstract}
The presence of dark energy in the Universe is inferred directly and indirectly from a large body 
of observational evidence. The simplest and most theoretically appealing possibility is the vacuum 
energy density (cosmological constant). However, although in agreement with current observations, 
such a possibility exacerbates the well known cosmological constant problem, requiring a natural 
explanation for its small, but nonzero, value. In this paper we focus our attention on another 
dark energy candidate, one arising from gravitational \emph{leakage} into extra dimensions. We 
investigate observational constraints from  current measurements of angular size of high-$z$ 
compact radio-sources on accelerated models based on this large scale modification of gravity. The 
predicted age of the Universe in the context of these models is briefly discussed. We argue that 
future observations will enable a more accurate test of these cosmologies and, possibly, show that 
such models constitute a viable possibility for the dark energy problem.
\end{abstract}

\pacs{98.80; 98.80.E; 04.50.+h; 95.35.+d}
\maketitle

\section{INTRODUCTION}
Recent results based on the magnitude-redshift relation for extragalactic type 
Ia supernovae (SNe Ia) suggest an accelerating universe dominated by some kind of negative-pressure 
dark component, the so-called \emph{quintessence} \cite{perlmutter,riess}. The existence of this 
dark energy has also been confirmed, independently of the 
SNe Ia analyses, by combining the latest galaxy clustering data with cosmic microwave background 
(CMB) measurements \cite{efs}. 
Together, these results seem to provide the remaining piece of information 
connecting the inflationary flatness prediction ($\Omega_{\rm{T}} = 1$) with astronomical 
observations. Such a state of affairs has  stimulated the interest for more general models 
containing 
an extra component describing this dark energy and, simultaneously, accounting for the present 
accelerated stage of the Universe. The simplest and most theoretically appealing possibility is the 
vacuum energy (cosmological constant). Because of their observational successes, flat models with a 
relic cosmological constant are considered nowadays our best description of the observed 
Universe. However, we face at least a serious  problem when one considers a nonzero vacuum energy: 
in order to dominate 
the dynamics of the Universe only at recent times, a very small value for the cosmological constant 
($\Lambda_{o}  \sim 10^{-56}\rm{cm}^{-2}$) is required from observations, while naive estimates 
based on quantum field theories are 50-120 orders of magnitude larger, thereby originating an 
extreme fine tunning problem \cite{weinberg,sahni} or making a complete 
cancellation (from an unknown physical mechanism) seem more plausible. 

If the cosmological constant is null, something else must be causing the Universe to speed up. 
Several possible dark energy candidates have been discussed in the literature, e.g., a vacuum 
decaying energy  density, or a time varying $\Lambda$-term \cite{ozer}, a relic scalar field 
\cite{peebles}, an extra component, the so-called ``X-matter", which is simply characterized by an 
equation of state $p_x=\omega\rho_{x}$, where $-1 \leq \omega < 0$ \cite{turner} and that includes, 
as a particular case, models with a cosmological constant ($\Lambda$CDM) or still models based on 
the framework of brane-induced gravity \cite{dvali,dvali1,deff,deff1,deffZ}. In this last case, the 
basic idea is that our 4-dimensional Universe would be a surface or a brane embedded into a higher 
dimensional bulk space-time to which gravity can spread. Despite the fact that there is no 
cosmological constant on the brane, such scenarios explain the observed acceleration of the 
Universe because the bulk gravity sees its own curvature term on the brane as a negative-pressure 
dark component and accelerates the Universe \cite{deff1}. 

Brane world cosmologies have been discussed in different contexts. For 
example, the issue related to the cosmological constant problem has been addressed \cite{ccp}as 
well as the evolution of cosmological perturbations in the gauge-invariant formalism \cite{brand}, 
cosmological phase transitions \cite{cpt}, inflationary solutions \cite{cpt1}, baryogenesis 
\cite{dvali99}, 
stochastic background of gravitational waves \cite{hogan1}, singularity, homogeneity, flatness and 
entropy problems \cite{aaa}, among others (see \cite{hogan} for a discussion on the different 
perspectives 
of brane world models). In the observational front, some analyses \cite{deffZ} have shown that such 
models are in agreement with the most recent cosmological observations (see, however, 
\cite{avelino,dnew}). In this case, constraints from SNe Ia + 
CMB data require a flat universe with $\Omega_{\rm{m}} = 0.3$ and $\Omega_{\rm{r}_c} = 0.12$, where 
$\Omega_{\rm{r}_c}$ is the density parameter associated with the crossover distance between the  
4-dimensional and 5-dimensional gravities (see \cite{deff1} for details).

In the present work we focus our attention on these kinds of cosmologies. Following \cite{deffZ}, 
we 
study models based on the framework of the brane-induced gravity of Dvali {\it et al.} 
\cite{dvali,dvali1} that have been recently proposed in Refs. \cite{deff,deff1}. We will also 
consider only the case in which the bulk space-time is 5-dimensional. Our main purpose is to 
investigate some observational consequences of such scenarios with emphasis on the constraints 
provided by observations of the angular size of high-$z$ milliarcsecond radio-sources. 

We structured this paper as follows. In Section II the basic field equations and distance formulas 
are presented. We also briefly discuss the predicted age of the Universe. In Section III we use 
measurements of the angular size of high-$z$ milliarcsecond radio sources \cite{gurv1} to constrain 
the 
free parameters of the model. We show that a good agreement between theory and observation is 
possible if $\Omega_{\rm{m}} \leq 0.38$,  $\Omega_{\rm{r}_c} \leq 0.29$ and $\Omega_{\rm{m}} \leq 
0.09$, $\Omega_{\rm{r}_c} \leq 0.29$ ($68\%$ c.l.) for values of the characteristic length of the 
sources between $l \simeq 20h^{-1} - 30h^{-1}$ pc, respectively. In particular we find that a 
slightly closed, accelerated  model with $\Omega_{\rm{m}} = 0.06$, $\Omega_{\rm{r}_c} = 0.28$, and 
$l = 27.06h^{-1}$ pc is the best fit for these data.

\section{Field equations, distance formulas and the age of the universe}

The geometry of our 4-dimensional Universe is described by the Friedmann-Robertson-Walker (FRW) 
line element ($c = 1$)
\begin{equation}
ds^2 = dt^2 - R^{2}(t) \left[{dr^{2} \over 1 - kr^{2}} + r^{2} (d
\theta^2 + \rm{sin}^{2} \theta d \phi^{2})\right],
\end{equation}
where $k = 0$, $\pm 1$ is the curvature parameter of the spatial section, $r$, $\theta$, and $\phi$ 
are dimensionless comoving coordinates, and $R(t)$ is the scale factor. The Friedmann's equation 
for the kind of models we are considering is \cite{deff1,deffZ}
\begin{equation} 
\left[\sqrt{\frac{\rho}{3M_{pl}^{2}} + \frac{1}{4r_{c}^{2}}} + \frac{1}{2r_{c}}\right]^{2} = H^{2} 
+ \frac{k}{R(t)^{2}},
\end{equation} 
where $\rho$ is the energy density of the cosmic fluid, $H$ is the Hubble parameter, $M_{pl}$ is 
the Planck mass, and $r_c = M_{pl}^{2}/2M_{5}^{3}$ is the crossover scale defining the 
gravitational interaction among particles located on the brane ($M_5$ is the 5-dimensional reduced 
Planck mass). For distances smaller than $r_c$ the force experienced by two punctual sources is the 
usual 4-dimensional gravitational $1/r^{2}$ force whereas for distances larger than $r_c$ the 
gravitational force follows the 5-dimensional $1/r^{3}$ behavior \cite{comm}.

Equation (2) implies that the normalization condition is given by
\begin{equation}
\Omega_k + \left[\sqrt{\Omega_{\rm{r_c}}} + \sqrt{\Omega_{\rm{r_c}} + \Omega_{\rm{m}}}\right]^{2} = 
1
\end{equation}
where $\Omega_{\rm{m}}$ and $\Omega_k$ are the matter and curvature density parameters, 
respectively and 
\begin{equation}
\Omega_{\rm{r_c}} = 1/4r_c^{2}H^{2},
\end{equation} 
is the density parameter associated to the crossover radius $r_c$.

The deceleration parameter, usually defined as $q_o = -R\ddot{R}/\dot{R}^{2}|_{t_{o}}$, now takes 
the following form
\begin{eqnarray}
q_o & = & \frac{3}{2}\Omega_{\rm{m}}\left[ 1 + \sqrt{\frac{\Omega_{\rm{r_c}}}{\Omega_{\rm{r_c}} + 
\Omega_{\rm{m}}}}\right] \\ \nonumber & & \quad    \quad    \quad    \quad    \quad    \quad    - 
\left[\sqrt{\Omega_{\rm{r_c}}} + \sqrt{\Omega_{\rm{r_c}} + \Omega_{\rm{m}}}\right]^{2} .
\end{eqnarray}
For $\Omega_k = 0$ (flat case), the above expression reduces to
\begin{equation}
q_o = \frac{3}{2}\Omega_{\rm{m}}\left[ 1 + \sqrt{\frac{\Omega_{\rm{r_c}}}{\Omega_{\rm{r_c}} + 
\Omega_{\rm{m}}}}\right] - 1.
\end{equation}

\begin{figure}
\vspace{.2in}
\centerline{\psfig{figure=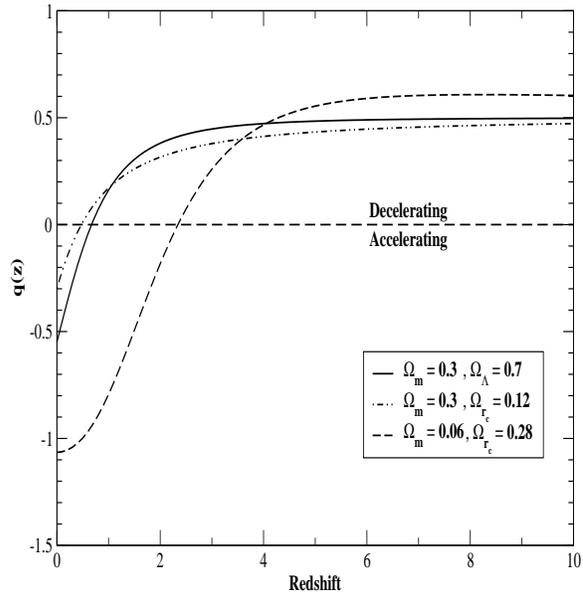,width=3.5truein,height=4.0truein,angle=-90}
\hskip 0.1in} 
\caption{Deceleration parameter as a function of redshift for some selected values of 
$\Omega_{\rm{m}}$ and $\Omega_{\rm{r_c}}$. The current best fit $\Lambda$CDM case is also shown 
for comparison. The horizontal line labeled decelerating/accelerating ($q_o = 0$) divides models 
with a decelerating or accelerating expansion at a given redshift.}
\end{figure}

Figure 1 shows the behavior of the deceleration parameter as a function of redshift for selected 
values of $\Omega_{\rm{m}}$ and $\Omega_{\rm{r_c}}$. As explained above, although there is no 
cosmological constant on the brane, brane-world models allows periods of accelerated expansion 
because 
the bulk gravity sees its own curvature term on the brane as a negative-pressure 
dark component \cite{deff1}. The best fit $\Lambda$CDM case is also showed for the sake of 
comparison. 
Note that at late times brane-cosmologies with $\Omega_{\rm{r_c}} = 0.3$ accelerates slower than 
$\Lambda$CDM models with $\Omega_{\Lambda} = 0.7$ and the same value of $\Omega_{\rm{m}}$. For the 
best fit model found in Ref. \cite{deffZ}, i.e., $\Omega_{\rm{m}} = 0.3$ and $\Omega_{\rm{r_c}} = 
0.12$, the accelerated expansion begins at $z \simeq 0.5$ whereas for $\Lambda$CDM we find $z 
\simeq 0.7$. For our best fit, found in Section 3, we see that the Universe always accelerates at a 
faster rate than the best fit $\Lambda$CDM model. In this case, the Universe begins to 
accelerate at $z \simeq 2.3$ (see also \cite{sahini} for a more detailed discussion on this topic).

From Eqs. (1) and (2), it is straightforward to show that the comoving distance $r(z)$ is given by
\begin{equation}  
r(z) = \frac{1}{R_o H_o |\Omega_k|^{1/2}}\sum\left[|\Omega_k|^{1/2} \int_{x'}^{1} {dx \over 
x^{2}f(\Omega_{j}, x)}\right], 
\end{equation} 
where the subscript $o$ denotes present day quantities, $x' = {R(t) \over R_o} = (1 + z)^{-1}$ is a 
convenient integration variable and the function $\sum(r)$ is defined by one of the following 
forms: $\sum(r) = \mbox{sinh}(r)$, $r$, and $\mbox{sin}(r)$, respectively, for open, flat and 
closed geometries. The dimensionless function $f(\Omega_{j}, x)$ takes the following form:
\begin{equation}
f(\Omega_{j}, x) = \left[\Omega_k x^{-2} + \left(\sqrt{\Omega_{\rm{r_c}}} + 
\sqrt{\Omega_{\rm{r_c}} 
+ \Omega_{\rm{m}}x^{-3}}\right)^{2}\right]^{1/2},
\end{equation}
where $j$ stands for $m$, $r_c$ and $k$. 

\begin{figure}
\vspace{.2in}
\centerline{\psfig{figure=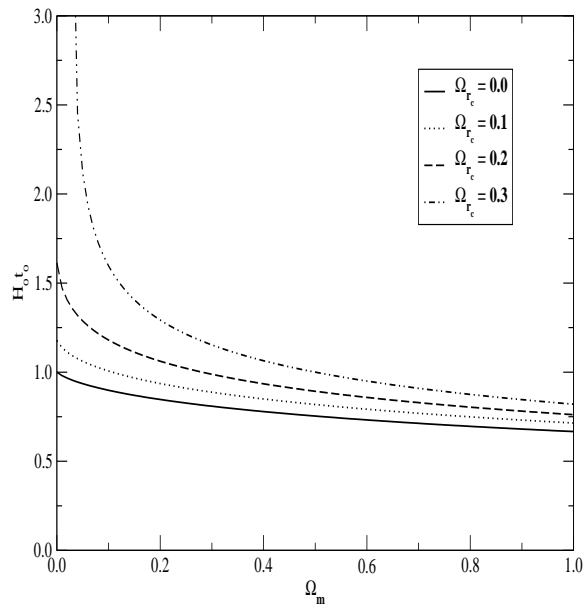,width=3.5truein,height=4.0truein,angle=-90}
\hskip 0.1in} 
\caption{Dimensionless age parameter as a function of $\Omega_{\rm{m}}$ for some selected 
values of 
$\Omega_{\rm{r_c}}$. As explained in the text, for a fixed value of $\Omega_{\rm{m}}$ the larger 
the contribution of $\Omega_{\rm{r_c}}$ the larger the age predicted by the model. }
\end{figure}

Similarly, the predicted age of the Universe as a function of the redshift can be written as
\begin{equation}  
t_z = H_o^{-1}\int_{0}^{x'} {dx \over x f(\Omega_{j}, x)}. 
\end{equation} 
As one may check from Eqs. (2), (5), (7) and (9), for $\Omega_{\rm{r_c}} = 0$, the standard 
relations are recovered. In Fig. 2 we show the dimensionless age parameter $H_ot_o$ as a function 
of $\Omega_{\rm{m}}$ for several values of $\Omega_{\rm{r_c}}$. Note that for a fixed value of 
$\Omega_{\rm{m}}$ the predicted age of the Universe is larger for larger values of 
$\Omega_{\rm{r_c}}$, 
thereby showing, similarly to what happens in the $\Lambda$CDM context, that the class of models 
studied here is  
efficient to solve the ``already" classical age of the Universe problem. For example, if 
$\Omega_{\rm{m}} = 0.3$, as sugested by dynamical estimates on scales up to about $2h^{-1}$ Mpc 
\cite{calb}, and $\Omega_{\rm{r_c}} = 0.15$ we find $t_o \simeq 13$ Gyr ($H_o = 
70{\rm{Km.s^{-1}.Mpc^{-1}}}$) or, in terms of the age parameter, $H_ot_o \simeq 0.93$, a value that 
is compatible with the most recent age estimates of globular clusters \cite{carreta,chab} as well 
as very close to some determinations based on SNe Ia data \cite{riess,tonry}.

\begin{figure} [t]
\vspace{.2in}
\centerline{\psfig{figure=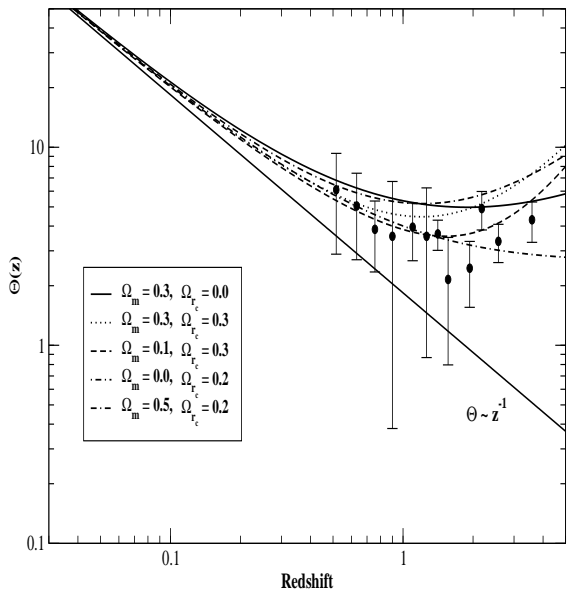,width=3.5truein,height=4.0truein,angle=-90}
\hskip 0.1in} 
\caption{Angular size vs redshift diagram for milliarcsecond radio sources data of Gurvits {\it et 
al.} [24]. The curves correspond to characteristic linear size $l = 27.06h^{-1}$ pc (D = 1.84 mas). 
Thick solid curve is the prediction of the standard open model $\Omega_{\rm{r_c}} = 0$.}
\end{figure}

\section{Constraints from high-$z$ angular size measurements}

In this section we study the constraints from angular size measurements of high-$z$ radio sources 
on the free parameters of the model. In the following we briefly outline our main assumptions for 
this analysis. Our approach is based on Ref. \cite{alcaniz}.

The angular size-redshift relation for a rod of characteristic length $l$ can 
be written as \cite{sand}
\begin{equation}
\theta (z) = \frac{D(1 + z)}{r(z)}.
\end{equation}
In the above expression $D = 100lh$ is the angular size scale expressed in milliarcsecond (mas) for 
$l$ measured in parsecs (compact sources).

In order to constrain the parameters $\Omega_{\rm{m}}$ and $\Omega_{\rm{r_c}}$ we use the angular 
size data for milliarcsecond radio sources recently compiled by Gurvits {\it et al.} \cite{gurv1}. 
This data set, originally composed by 330 sources distributed over a wide range of redshifts 
($0.011 \leq z \leq 4.72$), was reduced to 145 sources with spectral index $-0.38 \leq \alpha \leq 
0.18$ and total luminosity $Lh^{2} \geq 10^{26}$ W/Hz in order to minimize any possible dependence 
of angular size on spectral index and/or linear size on luminosity. This new subsample was 
distributed into 12 bins with 12-13 sources per bin. Two points, however, should be stressed before 
discussing the resulting diagrams. First of all, the determination of cosmological parameters is 
strongly dependent on the characteristic length $l$ (see, e.g., \cite{alcaniz}). In the absence of 
a statistical study describing the intrinsic length distribution of the sources, we follow 
\cite{alcaniz} and, instead of assuming a specific value for the mean projected linear 
size, we have worked on the interval $l \simeq 20h^{-1}  - 30h^{-1}$ pc, i.e., $l \sim O(40)$ pc 
for $h = 0.65$, or equivalently, $D = 1.4 - 2.0$ mas (see also \cite{gurv} for a detailed 
discussion on this topic). Second, following Kellermann \cite{kelle}, we assume 
that compact radio sources are free of the evolutionary and selection effects that have bedevilled 
attempts to use extended double radio  source in this context (see, for exemple, \cite{buc}), as 
they are deeply embedded in active galact nuclei, and, therefore, their morphology 
and kinematics do not depend considerably on the changes of the intergalactic medium. Moreover, 
these sources have typical ages of some tens of years, i.e., it is reasonable to suppose that a 
stable population is estabilished, characterized by parameters that do not change with the cosmic 
epoch \cite{jack}.

Following a procedure similar to that described in \cite{alcaniz}, we determine the 
cosmological parameters $\Omega_{\rm{m}}$ and $\Omega_{\rm{r_c}}$ through a $\chi^{2}$ minimization 
for a range of $\Omega_{\rm{m}}$ and $\Omega_{\rm{r_c}}$ spanning the interval [0, 1] in steps of 
0.02,
\begin{equation}
\chi^{2}(l, \Omega_{\rm{m}}, \Omega_{\rm{r_c}}) =
\sum_{i=1}^{12}{\frac{\left[\theta(z_{i}, l, \Omega_{\rm{m}}, \Omega_{\rm{r_c}}) - 
\theta_{oi}\right]^{2}}{\sigma_{i}^{2}}},
\end{equation}
where $\theta(z_{i}, l, \Omega_{\rm{m}}, \Omega_{\rm{r_c}})$ is given by Eqs. (7) and (10) and 
$\theta_{oi}$ is the observed values of the angular size with errors $\sigma_{i}$ of the $i$th bin 
in the sample.

\begin{figure} [t]
\vspace{.2in}
\centerline{\psfig{figure=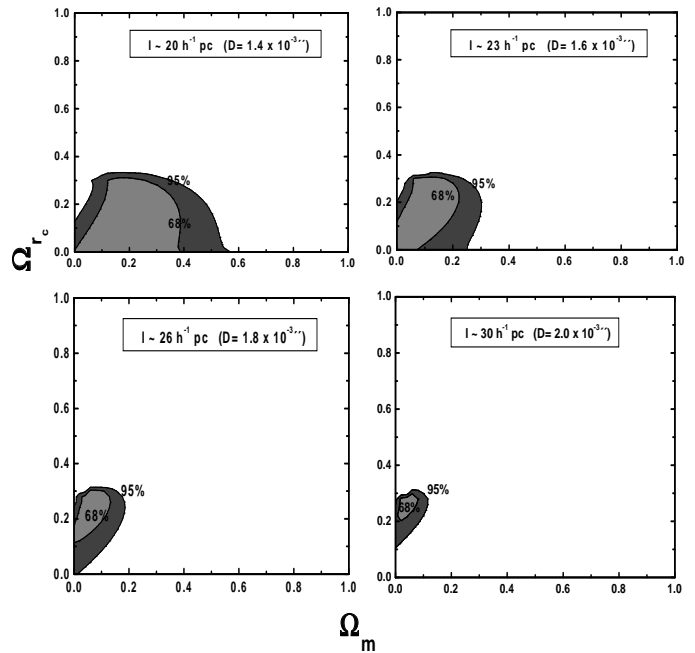,width=3.6truein,height=3.5truein,angle=0}
\hskip 0.1in} 
\caption{Confidence regions in the $\Omega_{\rm{m}} - \Omega_{\rm{r_c}}$ plane according to the 
updated sample of angular size data of Gurvits {\it et al.} [24]. Solid lines in each panel show 
contours of constant likelihood (95$\%$ and 68$\%$).}
\end{figure}

In Fig. 3 we show the binned data of the median angular size plotted as a function of redshift for 
several values of $\Omega_{\rm{m}}$ and $\Omega_{\rm{r_c}}$. For comparison we also show the 
standard prediction (thick line). Fig. 4 displays the $95\%$ and $68\%$ c.l. limits from angular 
size data on the $\Omega_{\rm{m}} - \Omega_{\rm{r_c}}$ plane for the interval $l \simeq 20h^{-1}  - 
30h^{-1}$ pc. Note that the limits on the plane are more restrictive for increasing values of the 
characteristic length $l$. It happens because for $z \sim 2$ (where most of our data points are 
concentrated) the parameter $\Omega_{\rm{r_c}}$ has a behavior similar to a cosmological constant 
or quintessence, i.e., it increases the distance between two different redshifts. In this way, 
according to Eq. (10), for the same $\theta_{oi}$'s the larger the value of $l$ the larger the 
value 
of $r(z)$ that is required or, equivalently, the smaller the value of $\Omega_{\rm{m}}$. For $l = 
20.58h^{-1}$ pc (D = 1.4 mas) the peak of likelihood is located at $\Omega_{\rm{m}} = 0.22$ and 
$\Omega_{\rm{r_c}} = 0.18$. This assumption provides $\Omega_{\rm{m}} \leq 0.38$ and 
$\Omega_{\rm{r_c}} \leq 0.29$ at 1$\sigma$. In the subsequent panels of the same figure similar 
analyses are displayed for $l = 23.53h^{-1}$ pc (D = 1.6 mas), $l = 26.47h^{-1}$ pc (D = 1.8 mas) 
and $l = 29.41h^{-1}$ pc (D = 2.0 mas). For an analysis independent of the choice of the 
characteristic length $l$, i.e., minimizing Eq. (11) for $\Omega_{\rm{m}}$, $\Omega_{\rm{r_c}}$ and 
$l$, we obtain $\Omega_{\rm{m}} = 0.06$, $\Omega_{\rm{r_c}} = 0.28$ and $l = 27.06h^{-1}$ pc (D 
= 1.84 mas) as the best fit for these data with $\chi^{2} = 4.25$ and 9 degrees of freedom. We also 
remark  that although not discussed here it is possible to determine the influence of 
$\Omega_{\rm{r_c}}$ on the critical redshift $z_m$ at which the angular size takes its minimal 
value. However as shown elsewhere \cite{jailS}, this test cannot discriminate among world models 
since different scenarios provide similar values of $z_m$.

An elementary combination of our best fit with Eq. (4) enables us to estimate $r_c$ (the crossover 
distance between 4-dimensional and 5-dimensional gravity) in terms of the Hubble radius $H_o^{-1}$. 
One obtains,
\begin{equation} 
r_c \simeq 0.94 H_o^{-1}.
\end{equation}
Such a value is slightly different from that one found by Deffayett {\it et al.} \cite{deff} using 
SNe Ia and CMB data. In their analysis it was found as a concordance model for these two tests a 
flat model with $\Omega_{\rm{m}} = 0.3$ and  $\Omega_{\rm{r_c}} = 0.122$, leading to an estimate of 
the crossover radius of $r_c \simeq 1.4H_o^{-1}$. Naturally, with new results from different 
cosmological tests it will be possible to delimit the $\Omega_{\rm{m}} - \Omega_{\rm{r_c}}$ plane 
more precisely. An analysis on the observational constraints from statistics of gravitational 
lenses will appear in a forthcoming communication \cite{alc}.

\section{conclusion}

Based on a large body of observational evidence, a consensus is beginning to emerge that we live 
in a flat, accelerated universe composed of $\sim$ 1/3 of matter (barionic + dark) and $\sim$ 2/3 
of a negative-pressure dark component. However, since the nature of this dark energy is still not 
well understood, an important task nowadays in cosmology is to investigate the existing 
possibilities in the light of the current observational data. In this paper we have focused our 
attention on some observational aspects of brane world cosmologies. These models, inspired on 
superstring-M theory, explain the observed acceleration of the Universe through a large scale 
modification of gravity arising from a gravitational leakage into an extra dimension \cite{deffZ}. 
We showed that their predicted age of the Universe is compatible with the most recent age estimates 
of globular clusters and, therefore, that there is no age crisis in the context of these models. By 
using a large sample of milliarcsecond radio sources recently updated and extended by Gurvits {\it 
et al.} \cite{gurv1} we obtained, as the best fit for these data, a slightly closed, accelerated  
universe with $\Omega_{\rm{m}} = 0.06$ and $\Omega_{\rm{r_c}} = 0.28$. Such values lead to an 
estimate of the crossover radius of $r_c \simeq 0.94 H_o^{-1}$.

\acknowledgments
It is my pleasure to acknowledge helpful discussions with Professor J. A. S. Lima and Professor C. 
J. Hogan. I also thank Professor L. I. Gurvits for sending his compilation of the data and J. V. 
Cunha 
for producing Figure 4. This research is supported by the Conselho Nacional de Desenvolvimento 
Cient\'{\i}fico e  Tecnol\'{o}gico (CNPq - Brazil) and CNPq(620053/01-1-PADCT III/Milenio)


\begin{references}
\bibitem{comm} As is well known, according to Gauss' law the gravitational force falls off as $r^{1 
- S}$ where $S$ is the number of spatial dimensions.
\end{references}
\end{document}